\documentclass[prl,twocolumn,superscriptaddress]{revtex4}
\usepackage{graphicx}
\usepackage{dcolumn}
\usepackage{amsmath,amsthm,amssymb}
\usepackage{subfigure}

\def\ep{\varepsilon}

\begin{document}
\title{A comment on AdS collapse of a scalar field in higher dimensions}

\author{Joanna Ja\l{}mu\.zna}
\affiliation{Institute of Mathematics, Jagiellonian
University, Krak\'ow, Poland}
\author{Andrzej Rostworowski}
\affiliation{Institute of Physics, Jagiellonian
University, Krak\'ow, Poland}
\author{Piotr Bizo\'n}
\affiliation{Institute of Physics, Jagiellonian
University, Krak\'ow, Poland}
\date{\today}
\begin{abstract}
We point out that the weakly turbulent instability of anti-de Sitter space, recently found in \cite{br} for four dimensional spherically symmetric Einstein-massless-scalar field equations with negative cosmological constant, is present in all dimensions $d+1$ for $d\geq 3$, contrary to a claim made in \cite{gl}.
\end{abstract}

\maketitle

In a recent paper \cite{br} two of us reported on numerical simulations which indicate that anti-de Sitter (AdS) space is unstable against the formation of a black hole under arbitrarily small generic perturbations.  This instability was conjectured to be triggered by a resonant mode mixing  which moves energy from low to high frequencies.
Although the simulations of \cite{br} were done only in four spacetime dimensions, it was clear from the nonlinear perturbation analysis given there that the same weakly turbulent mechanism of generating instability operates in all supercritical dimensions $d+1$ for $d\geq 3$. Thus, we found it surprising that Garfinkle and Pando Zayas, who later looked at the same problem in $4+1$ dimensions (apparently unaware of Ref.\cite{br} as they did not cite it), wrote in \cite{gl}: "We [...] establish that for small values of the initial amplitude of the scalar field there is no [sic!] black hole formation, rather, the scalar field performs an oscillatory motion typical of geodesics in AdS." The purpose of this comment is to show that, contrary to the above quoted claim, $AdS_{d+1}$ is unstable against gravitational collapse for all $d\geq 3$.  To this end, we first generalize the formalism of \cite{br} to higher dimensions and  recall the key argument for collapse of arbitrarily small generic initial data. Second,  using our code in $d=4$ we evolve numerically the same initial data as those of \cite{gl} and show that, as expected, they collapse. Finally, we try to identify a possible source of the error in \cite{gl}.

 We parametrize the $(d+1)$--dimensional  asymptotically AdS metric by the ansatz
\begin{equation}
\label{adsd+1:ansatz}
ds^2\! =\! \frac {\ell^2}{\cos^2{\!x}}\left( -A e^{-2 \delta} dt^2 + A^{-1} dx^2 + \sin^2{\!x} \,  d\Omega^2_{d-1}\right)\,,
\end{equation}
where $\ell^2=-d(d-1)/(2\Lambda)$, $d\Omega^2_{d-1}$ is the round metric on $S^{d-1}$, $-\infty<t<\infty$, $0\leq x<\pi/2$, and  $A$, $\delta$ are functions of $(t,x)$. For this ansatz the evolution of a self-gravitating massless scalar field $\phi(t,x)$ is governed by the following system (using units in which $8\pi G=d-1$)
\begin{equation}
\label{ms_in_ads_d+1:eq_wave}
\dot\Phi = \left( A e^{-\delta} \Pi \right)', \quad \dot \Pi = \frac{1}{\tan^{d-1}{\!x}}\left(\tan^{d-1}{\!x} \,A e^{-\delta} \Phi \right)',
\end{equation}
\begin{align}
\label{ms_in_ads_d+1:eq_00}
A' \!&= \!\frac{d-2+2\sin^2{\!x}} {\sin{x}\cos{x}} \, (1-A) - \sin{x}\cos{x} \, A \left( \Phi^2 + \Pi^2 \right),
\\
\label{ms_in_ads_d+1:eqs_10_11}
\delta' \!&=\! -  \sin{x}\cos{x} \left( \Phi^2 + \Pi^2 \right),
\end{align}
where ${}^{\cdot}=\partial_t$, ${}'=\partial_x$,
$\Phi= \phi'$ and $\Pi= A^{-1} e^{\delta} \dot \phi$.
 We want to solve the system (2-4) for small smooth perturbations of  AdS solution $\phi=0, A=1, \delta=0$. Smoothness at the center implies that near $x=0$
\begin{align}\label{x=0}
    \phi(t,x)&= f_0(t)+\mathcal{O}(x^2),\quad  \delta(t,x)= \mathcal{O}(x^2), \nonumber \\
    A(t,x)&=1+\mathcal{O}(x^2)\,,
\end{align}
where we used normalization $\delta(t,0)=0$ so that $t$ is the proper time at the center.
Smoothness at spatial infinity and finiteness of the total mass $M$ imply that near $x=\pi/2$ we must have (using $\rho=\pi/2-x$)
\begin{align}\label{pi2}
    \phi(t,x)&= f_{\infty}(t)\, \rho^d+\mathcal{O}\left(\rho^{d+2}\right),\quad
    \delta(t,x)=\delta_{\infty}(t)+ \mathcal{O}\left(\rho^{2d}\right), \nonumber \\
    A(t,x)&=1- M \rho^d+\mathcal{O}\left(\rho^{d+2}\right),
\end{align}
where the power series expansions are uniquely determined by $M$  and  the functions $f_{\infty}(t)$, $\delta_{\infty}(t)$ (which in turn are determined by the evolution of initial data).
One can show that the initial-boundary value problem for the system (2-4) together with the regularity conditions (5) and (6) is locally well-posed.

In \cite{br} the instability of $AdS_4$ was conjectured to result from the resonant mode mixing which moves energy from low to high frequencies. It was argued that this process of energy concentration on increasingly small spatial scales must be eventually cut off by the formation of a black hole. It is easy to see that the same mechanism is at work for all $d \ge 3$. This follows from the fact that, using the PDE terminology, the system (2-4) is fully resonant. More precisely, the spectrum of  the linear self-adjoint operator, which governs the evolution of linearized perturbations of $AdS_{d+1}$,
$
L = - \tan^{1-d}{x}\, \partial_x \left( \tan^{d-1}{x}\, \partial_x \right)\,,
$
is given by $\omega_j^2=(d+2j)^2$, ($j=0,1,...$). The key point is that the frequencies $\omega_j$ are equally spaced, so already at the third order of nonlinear perturbation analysis one gets resonant terms for any frequency $\omega_j$ such that $j = j_1 + j_2 - j_3$, where $j_k$ are indices of eigenmodes present in the initial data. Some of these resonances lead to secular terms which signal the onset of instability at time $t=\mathcal{O}(\varepsilon^{-2})$, where $\varepsilon$ measures the size of initial data.

To substantiate this heuristic argument we solve the system (\ref{ms_in_ads_d+1:eq_wave}-\ref{ms_in_ads_d+1:eqs_10_11}) in $d=4$ numerically, using the method of \cite{br}. For easy comparison of results, we take the same approximately ingoing Gaussian initial data as Ref.\cite{gl}: $\Phi(0,x)=\partial_x \phi(0,x)=\Pi(0,x)$, where
\begin{equation}\label{datainit}
\phi(0,x)=\frac{\ep}{\sqrt{3}} \exp\left(-\frac{(\tan{x}-r_0)^2}{\sigma^2}\right)
\end{equation}
with $r_0=4$, $\sigma=1.5$. As in \cite{gl} we set the AdS radius $\ell=1$, hence their and our  radial coordinates are related by $r=\tan{x}$, while the time coordinates are identical. We denote their amplitude $A$ by $\ep$ (to avoid conflict of notation); the factor $1/\sqrt{3}$ comes from the difference in units: we use $8 \pi G = 3$, while in \cite{gl} $8 \pi G =1$.
  Note that the data \eqref{datainit} slightly violate the regularity condition \eqref{x=0} since
  $\Phi(0,0)$ is not exactly zero, however an error generated by this 'corner singularity' is negligible.
\begin{figure}[h]
  \includegraphics[width=0.48\textwidth]{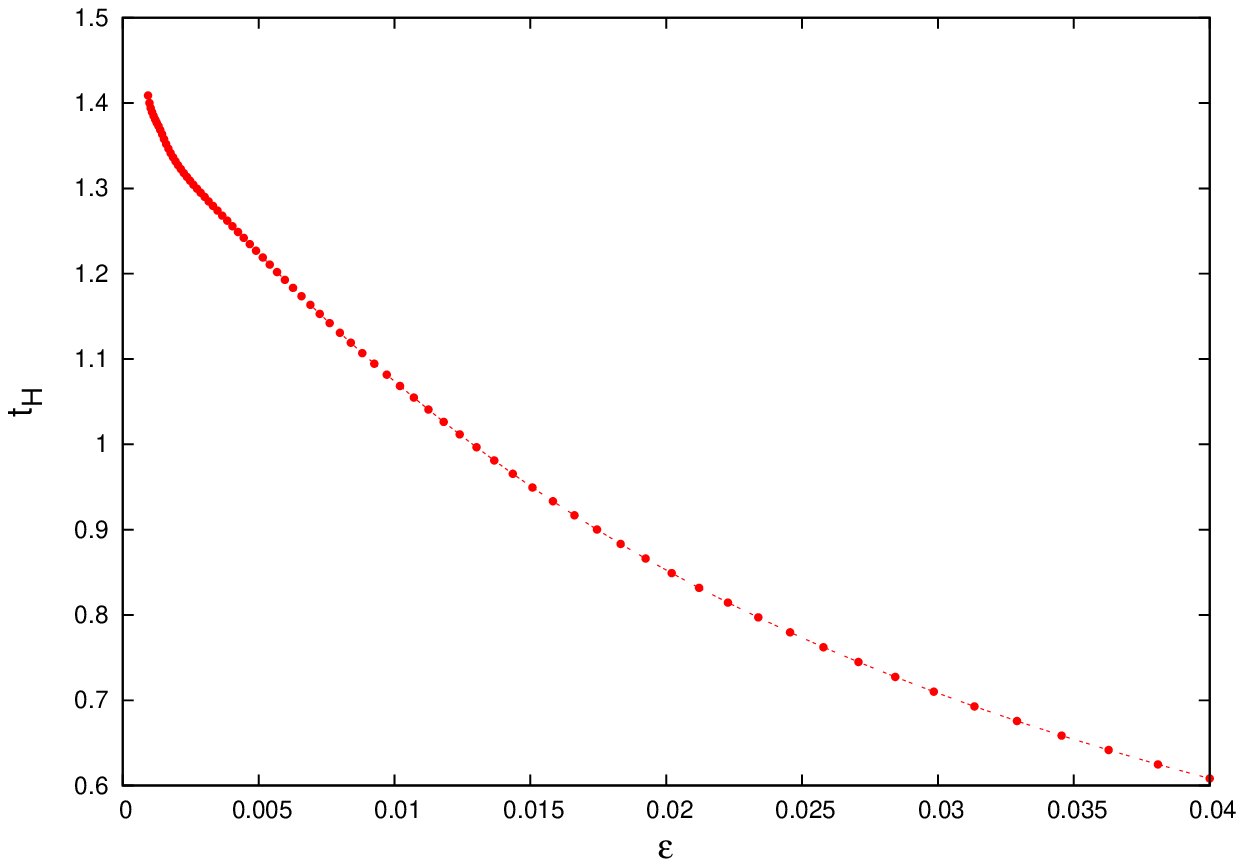}
  \includegraphics[width=0.48\textwidth]{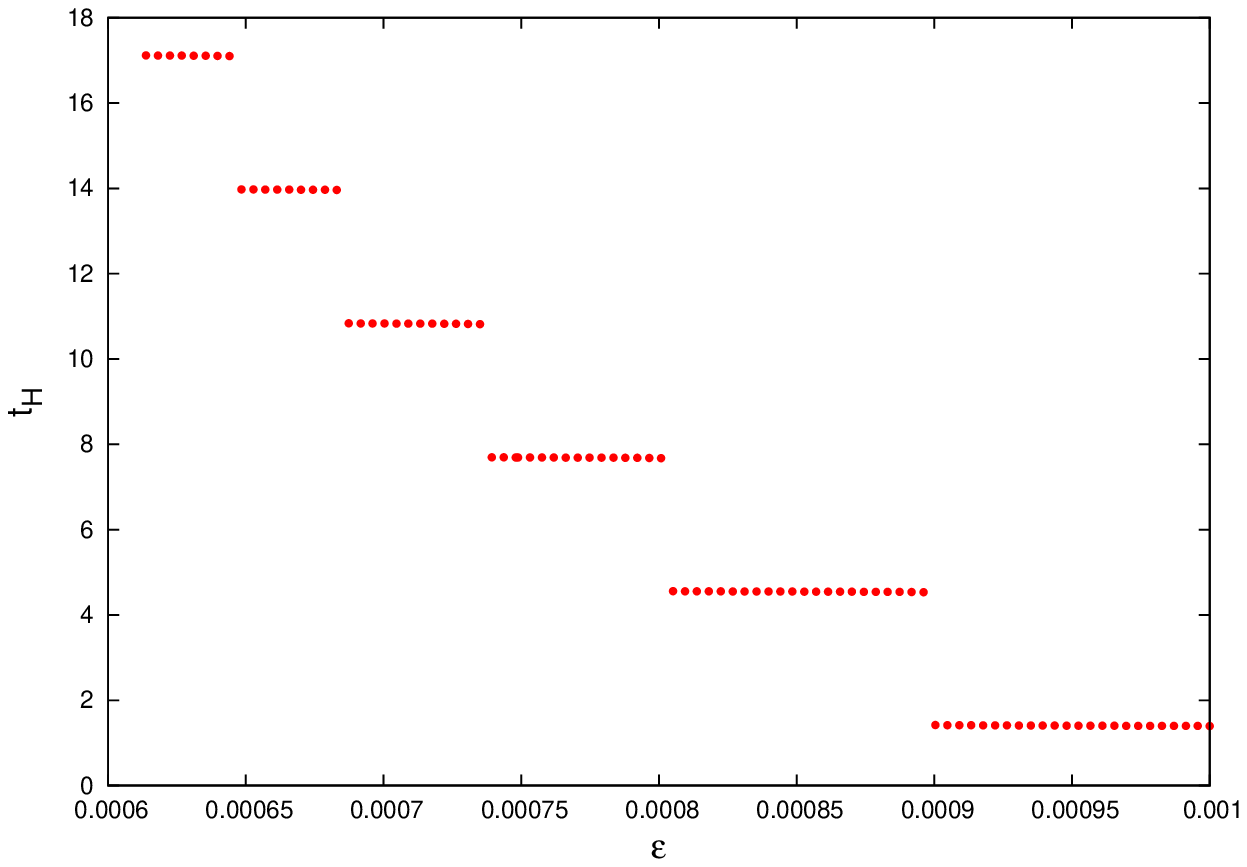}
  \caption{
  Time of horizon formation $t_H$ vs amplitude for initial data \eqref{datainit}. The upper plot
   depicts the first 'step' of the 'staircase' function $t_H(\ep)$ corresponding to large data solutions which collapse during the first implosion. The horizon radius varies from $x_H=0.5$ to zero (from right to left). For $\ep>0.005$, the plot
   coincides with Fig.~4 of \cite{gl} which verifies that our results agree with those of \cite{gl} for short enough times. The lower plot depicts a few further 'steps' of $t_H(\ep)$ corresponding to solutions which bounce several times from the AdS boundary before collapsing. }
  \label{fig1}
\end{figure}

For large $\ep$, the solution quickly collapses (the formation of an apparent horizon is detected by the metric function $A(t,x)$ touching zero at some $x_H$). As $\ep$ decreases,
the horizon radius takes the form of the right continuous sawtooth curve $x_H(\ep)$ (see Fig.~1 of \cite{br}) with jumps at critical points $\ep_n$ where $\lim_{\ep\rightarrow \ep_n^+} x_H(\ep)=0$
(the index $n$ counts the number of reflections from the AdS boundary before collapse). Accordingly, the time of horizon formation $t_H(\ep)$ is a monotone decreasing piecewise continuous function with jumps at each $\ep_n$ (see Fig.~1).

For small initial data the weakly nonlinear perturbation analysis described in \cite{br} predicts the onset of instability at time $t\sim \ep^{-2}$. Numerics indicates that for sufficiently small $\ep$ this scaling holds approximately almost all the way to the collapse, that is $t_H(\ep)\sim \ep^{-2}$. The  evidence for this fact is shown in Fig.~2 which depicts the evolution of three solutions with  small amplitudes differing by a factor of $\sqrt{2}$.
Note that this scaling implies that the computational cost of numerical evolution increases rapidly as $\ep$ decreases (since solutions have to be evolved for longer and longer times on finer and finer grids).

\begin{figure}[t]
  \includegraphics[width=0.48\textwidth]{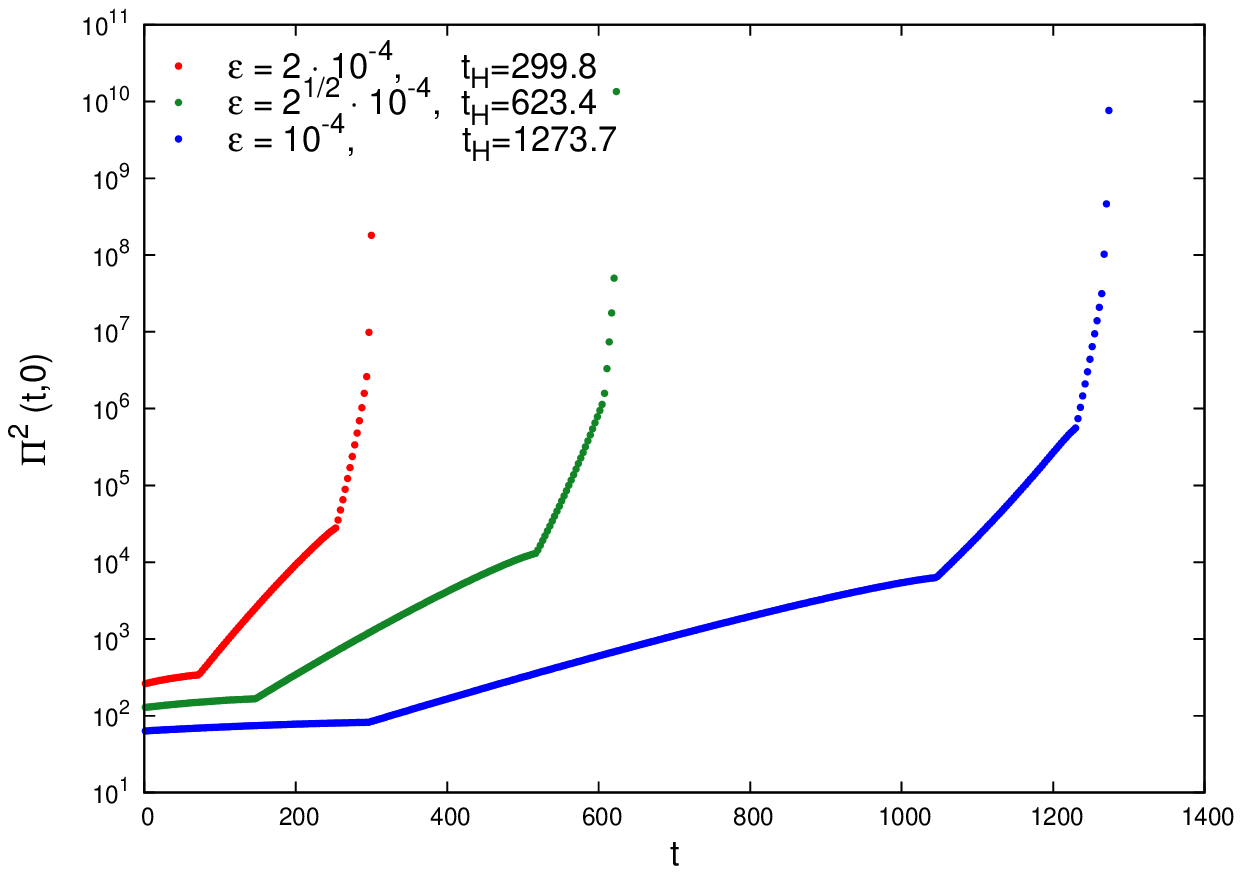}
   \includegraphics[width=0.48\textwidth]{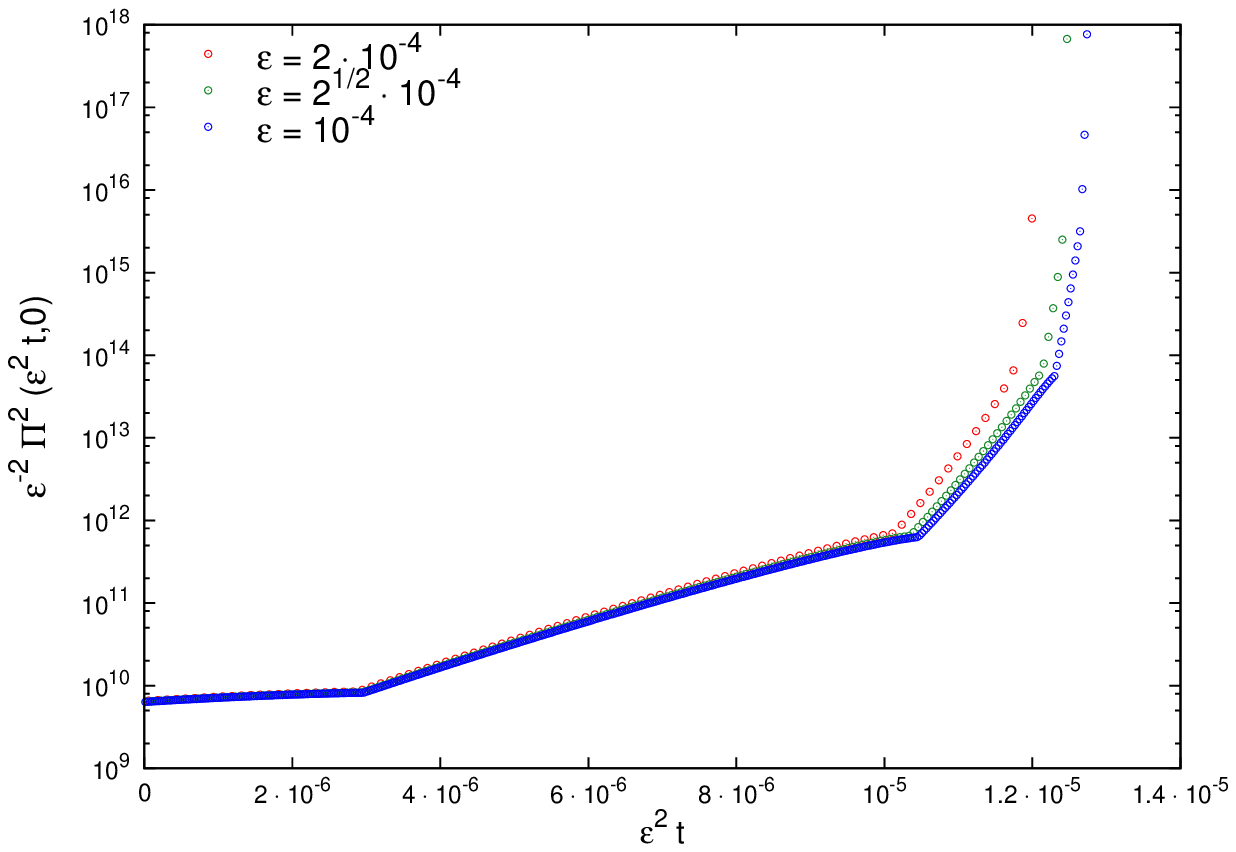}
   \caption{Upper plot: the upper envelope of $\Pi^2(t,0)$ for initial data \eqref{datainit} with three relatively small amplitudes. After making $95$ (for $\ep=0.0002$), $198$ (for $\ep=\sqrt{2}\cdot 0.0001$, and $405$ (for $\ep=0.0001$) reflections, all solutions eventually collapse. Lower plot:  the curves from the upper plot  after rescaling $\ep^{-2} \Pi^2(\ep^2 t,0)$ seem to converge to a limiting curve.}
  \label{fig2}
\end{figure}

 Finally, let us make a few remarks about the
 paper \cite{gl}. The content of that paper, when stripped of the standard promotional material for the application of AdS/CFT correspondence to the quark-gluon plasma, boils down to two statements concerning the solutions of five dimensional spherically symmetric Einstein-massless-scalar field equations with negative cosmological constant: (i) large initial data lead to collapse and the larger the data, the shorter the time of horizon formation; (ii) small initial data do not form black holes. The statement (i) is true but trivial. The statement (ii), as we have shown above, is false. We wonder what led the authors of \cite{gl} to reach this conclusion. The numerical method used in \cite{gl} is rather crude; it is based on the second order finite difference code on a fixed nonuniform grid. The radial coordinate $r$ is not compactified and the AdS timelike boundary at $r=\infty$ is mimicked by an artificial reflecting mirror at $r_{max}=10$. Although this fact might make  long-time quantitative results somewhat inaccurate (since the pulse gets reflected before reaching the true boundary), we do not think it had any effect on the claim (ii).
   We think that the culprit was the lack of sufficient resolution. In the first (v1) version of \cite{gl}, Fig.~2v1 shows three implosions for the solution with amplitude $\ep=0.001$, which (together with the discussion in the text) seems to suggest implicitly that no black hole forms later and "the scalar field performs an oscillatory motion" forever. We evolved these data with our code and observed horizon formation at $x_H=1.88\cdot 10^{-2}$ during the first implosion at time $t_H=1.398$ which means that the absence of collapse and the "oscillatory motion" of Fig.~2v1  are numerical artifacts. It appears that the authors of \cite{gl} have realized that their numerical simulations suffered from insufficient spatial resolution because in the second (v2) version of \cite{gl} the number of grid points was increased from 800 to 6400.
 We gather that this improvement of resolution has helped  to detect the collapse in Fig.2v1 during the first implosion because the new Fig.~2v2 depicts four implosions of a 'non-collapsing' solution with five times smaller amplitude $\ep=0.0002$.
  But this has just postponed the problem of the loss of resolution until a later time (not shown in the plot). Indeed, our numerical evolution of these data (shown in Fig.~2) yields horizon formation at $x_H\approx 4.8\cdot 10^{-4}$ after time $t_H\approx 299.8$. Even if the numerical simulation used to produce Fig.2v2 had been run  for such a long time, it would not have had enough resolution to capture the spatio-temporal structure of horizon formation  since this structure develops below  the first point of their grid. For comparison,
 our fourth-order code with global adaptive mesh refinement reached the level of $2^{17}+1$ grid points just before collapse. Anyway, the lesson is that the long-time
 numerical simulations of asymptotically AdS spacetimes are challenging even in spherical symmetry and
 one should be careful in jumping to conclusions about the  late time dynamics, especially without an analytic understanding of the problem.

\vskip 0.1cm \noindent \emph{Acknowledgments:}
JJ and PB thank David Garfinkle for the discussion. This work was supported in part by the Foundation for Polish Science under the MPD Programme "Geometry and Topology in Physical Models"
and by the NCN grant NN202 030740. The computations were performed on the Deszno supercomputer at the Institute of Physics, Jagiellonian University and at the Academic Computer Centre Cyfronet AGH using the PL-Grid infrastructure.

\end{document}